\documentclass[aps,prl,superscriptaddress,reprint,floats,citeautoscript,floatfix,nobibnotes,nofootinbib]{revtex4-1}

\usepackage[intlimits,tbtags]{amsmath}
\usepackage{times}
\usepackage{color}
\usepackage{amssymb}
\usepackage{graphicx}
\usepackage{bm}
\usepackage{array}
\usepackage{dcolumn}
\usepackage{bm}
\usepackage{setspace}

\newcolumntype{d}[1]{D{.}{.}{#1} }
\usepackage{xcolor}

\newcommand{\Tc}{$T_\text{c}$}

\newcommand{\jsnu}{Laboratory of Quantum Functional Materials Design and Application, School of Physics and Electronic Engineering, Jiangsu Normal University, Xuzhou 221116, China}

\newcommand{\jena}{Institut f\"ur Festk\"orpertheorie und -optik,
  Friedrich-Schiller-Universit\"at Jena, Max-Wien-Platz 1, 07743 Jena, Germany}
\newcommand{\halle}{Institut f\"ur Physik, Martin-Luther-Universit\"at
  Halle-Wittenberg, D-06099 Halle, Germany}

\newcommand{\poland}{Institute of Physics, Cz\c{e}stochowa University of Technology, Ave. Armii Krajowej 19, 42-200 Cz\c{e}stochowa, Poland}
\begin{document}
\title{Prediction of high-\Tc\ superconductivity in ternary actinium beryllium hydrides at low pressure}

\author{Kun Gao}
\affiliation{\jsnu}
\affiliation{\jena}
\author{Wenwen Cui}\email{wenwencui@jsnu.edu.cn}
\author{Jingming Shi}
\affiliation{\jsnu}
\author{Artur P. Durajski}
\affiliation{\poland}
\author{Jian Hao}\email{jian\_hao@jsnu.edu.cn}
\affiliation{\jsnu}
\author{Silvana Botti}\email{silvana.botti@uni-jena.de}
\affiliation{\jena}
\author{Miguel A. L. Marques}
\affiliation{\halle}
\author{Yinwei Li}
\affiliation{\jsnu}

\text{}

\begin{abstract}
Hydrogen-rich superconductors are promising candidates to achieve room-temperature superconductivity. However, the extreme pressures needed to stabilize these structures significantly limit their practical applications.
An effective strategy to reduce the external pressure is to add a light element M that binds with H to form MH$_x$ units, acting as a chemical precompressor. We exemplify this idea by performing \textit{ab initio} calculations of the Ac--Be--H phase diagram, proving that the metallization pressure of Ac--H binaries, for which critical temperatures as high as 200~K were predicted at 200~GPa, can be significantly reduced via beryllium incorporation. We identify three thermodynamically stable (AcBe$_2$H$_{10}$, AcBeH$_8$, and AcBe$_2$H$_{14}$) and four metastable compounds (\textit{fcc} AcBeH$_8$, AcBeH$_{10}$, AcBeH$_{12}$ and AcBe$_2$H$_{16}$). All of them are superconductors. In particular, \textit{fcc} AcBeH$_{8}$ remains dynamically stable down to 10\,GPa, where it exhibits a superconducting-transition temperature \Tc\ of 181~K. The Be--H bonds are responsible for the exceptional properties of these ternary compounds and allow them to remain dynamically stable close to ambient pressure. Our results suggest that high-\Tc\ superconductivity in hydrides is achievable at low pressure and may stimulate experimental synthesis of ternary hydrides.

\end{abstract}
\maketitle

\section{Introduction}

The renaissance of high-\Tc\ hydride superconductors began in 2004~\cite{li2014metallization,duan2014pressure} and has recently attracted growing attention, after a serious of spectacular experimental confirmations, such as superconductivity at 200~K in H$_3$S when subjected to an external pressure of 155~GPa~\cite{li2014metallization,duan2014pressure,drozdov2015conventional}, at 260~K and 170--180~GPa in LaH$_{10}$~\cite{liu2017potential,kostrzewa2020lah,peng2017hydrogen,drozdov2019superconductivity,somayazulu2019evidence}, at 215~K and 160~GPa or 172~GPa in CaH$_6$~\cite{wang2012superconductive,ma2022high,li2022superconductivity1}, at 220~K and 166~GPa or 237~GPa in YH$_6$~\cite{li2015pressure,kong2021superconductivity,troyan2019anomalous}, and at 243--262~K and 182--201~GPa in YH$_9$~\cite{kong2021superconductivity,snider2020superconductivity}. Compared to pure hydrogen, the pressures required to realize metallization in binary hydrides  are reduced considerably, but they are still so large to represent a serious limitation to any practical application. The search for high-\Tc\ superconductors at lower or even ambient pressure remains therefore an open challenge.

Thorough theoretical investigation of binary hydrides has revealed that most systems that display high-\Tc\ ($>$ 150 K) superconductivity are stable only above 150~GPa~\cite{kong2021superconductivity,troyan2019anomalous,drozdov2019superconductivity,somayazulu2019evidence}, while those that could be stabilized at lower pressure exhibit poor superconductivity~\cite{shimizu2002superconductivity,wang2021low,dietrich1974pressure,schirber1974concentration}. These relatively disappointing conclusions have however cleared the way for testing the largely unexplored family of ternary hydrides. In fact the additional chemical degree of freedom allows to enlarge enormously the search space, leading to exciting predictions, such as Li$_2$MgH$_{16}$ (\Tc\ = 473~K at 250~GPa)~\cite{sun2019route}, YCeH$_{20}$ (\Tc\ = 246~K at 350~GPa), LaCeH$_{20}$ (\Tc\ = 233~K at 250~ GPa)~\cite{song2022potential}, (La,Y)H$_{10}$ (\Tc\ = 253~K at 183~GPa)~\cite{semenok2021superconductivity}, and CaBeH$_8$ (\Tc\ = 254~K at 210~GPa)~\cite{zhang2022design}. Encouragingly, several high-\Tc\ ternary hydrides have already been experimentally synthesized, including (La,Ce)H$_9$ (\Tc\ = 48--172~K at 92--172~GPa)~\cite{bi2022giant,chen2022enhancement}, (La,Y)H$_{10}$ (\Tc\ = 253~K at 183~GPa)~\cite{semenok2021superconductivity}, and (La, Nd)H$_{10}$ (\Tc\ = 148~K at 180~GPa)~\cite{semenok2022effect}.

More specifically, we want to consider here the effect of adding a light element M that can bind with the H atoms to form small MH$_x$ units. Such units can serve as pre-compression factor and act on the lattice of the parent binary to potentially reduce further the metallization pressure. For example, it was shown that introducing C atoms in a S-H system to obtain CH$_4$ molecules yields dynamically stable ternary compounds with good superconducting properties, comparable with those of H$_3$S~\cite{cui2020route,sun2020computational}. Moreover, the incorporation of B or Be in lanthanum hydrides, forming Be/BH$_8$ units, makes this system dynamically stable down to below 50~GPa with a \Tc\ above 100~K~\cite{zhang2022design,liang2021prediction}. The formation of SiH$_8$ molecules with the same symmetry ensures the dynamical stability of BaSiH$_8$ and SrSiH$_8$ down to 3~GPa and 27~GPa, respectively~\cite{roman2022silico}. Moreover, BH$_4$ molecules intercalating \textit{fcc} lattices of alkaline metals X (X = K, Rb, Cs) lead to energetically stable, superconducting ternary hydrides XB$_2$H$_8$ with \Tc\ above 100~K at 10~GPa~\cite{li2022superconductivity,gao2021phonon}.

Particularly interesting are the results obtained by incorporating Be atoms in binary hydrides: in fact Be can act as electron donor to break the H$_2$ molecule and improve the superconducting properties, in the same way as Mg in Li$_2$MgH$_{16}$~\cite{sun2019route}. Moreover, it can lead to the formation of BeH$_8$ units that have the potential to reduce the metallization pressure of the parent hydride, as observed in LaBeH$_8$ (\Tc\ = 183~K at 20~GPa) and YBeH$_8$ (\Tc\ = 249~K at 100~GPa)~\cite{zhang2022design}. The light Be atoms bond with H, replacing some H--H bonds in the hydrides, which may lead to a reduction of the stabilization pressure. Additionally, doping with light elements can increase the average phonon frequencies, and if these vibrational modes are properly coupled with electrons at the Fermi energy ($E$$_F$), this can lead to an increase of \Tc. Therefore, we consider very promising to insert Be atoms in high-\Tc\ superconducting binary hydrides that are stable only at high pressures.

\begin{figure}[htp]
\centering
  \includegraphics[width=0.85\linewidth,angle=0]{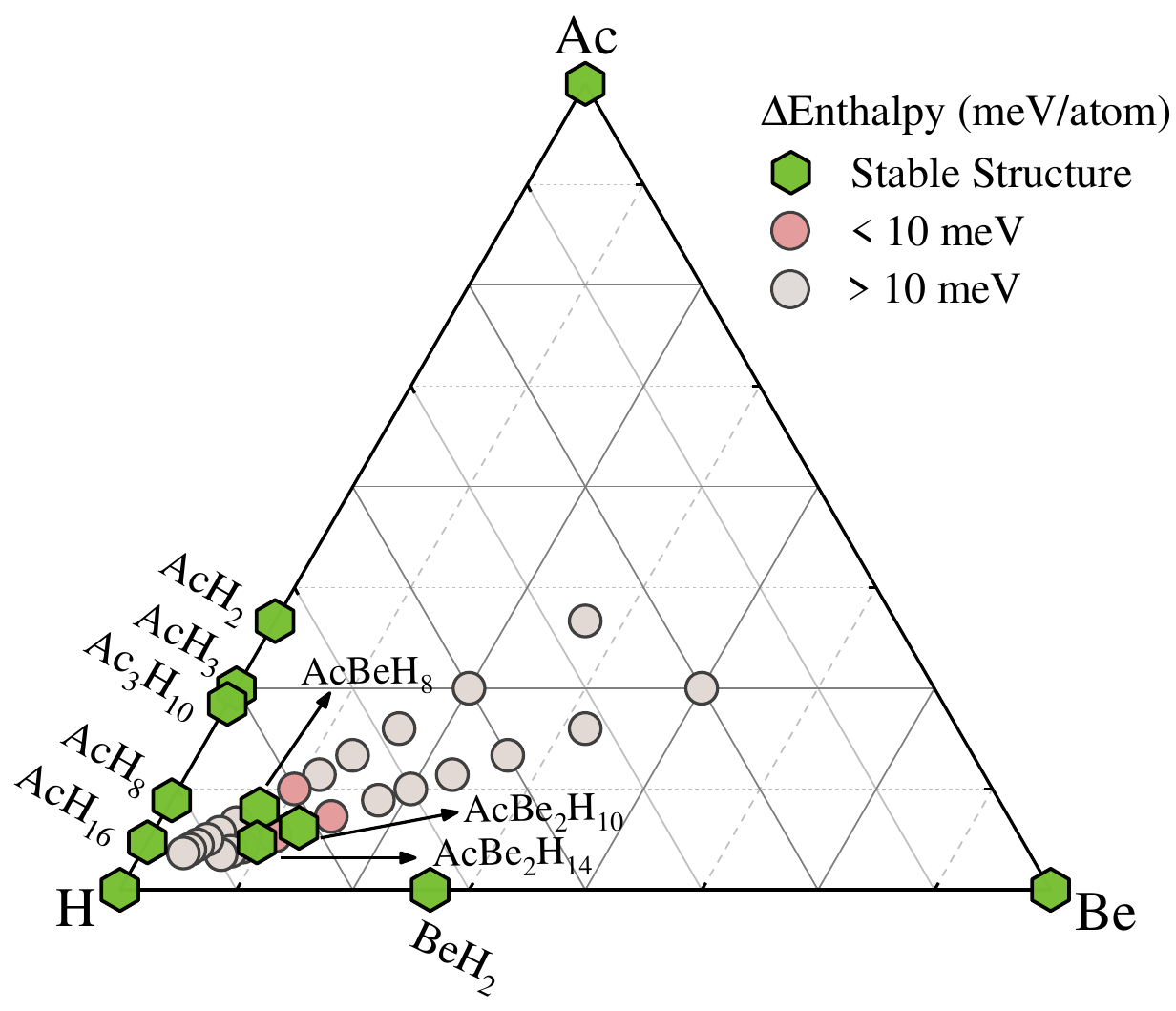}
  \caption{\label{fig:1} Thermodynamical stability. Ternary convex hull of Ac--Be--H at 200~GPa. The stable and metastable phases are depicted as hexagons (green) and circles (pink and gray), respectively.}
\end{figure}

Actinium hydrides have been thoroughly studied at high pressures~\cite{semenok2018actinium,yu2014exploration} and  several thermodynamically stable compounds have been proposed (e.g., AcH$_2$, AcH$_3$, Ac$_3$H$_{10}$, AcH$_8$, AcH$_{10}$ and AcH$_{16}$). Among these, AcH$_{16}$ is predicted to be a good superconductor with \Tc\ of 241~K at 150~GPa~\cite{semenok2018actinium}. Here we study the phase diagram of Ac--Be--H to find if the formation of BeH$_x$ units can act as chemical pre-compressor of the \textit{fcc} framework composed of the large Ac atoms, leading to ternary compounds that can be stable at low pressure. We perform therefore \textit{ab initio} structural prediction calculations of AcBe$_x$H$_y$ ($x$ = 1-2, $y$ = 1-20) in search of thermodynamically stable and metastable structures that are metallic and potentially superconducting. Successive electron-phonon coupling calculations are employed to evaluate the transition temperature for phonon-mediated superconductivity at different pressures.

\begin{figure*}[htp]
\centering
  \includegraphics[width=0.8\linewidth,angle=0]{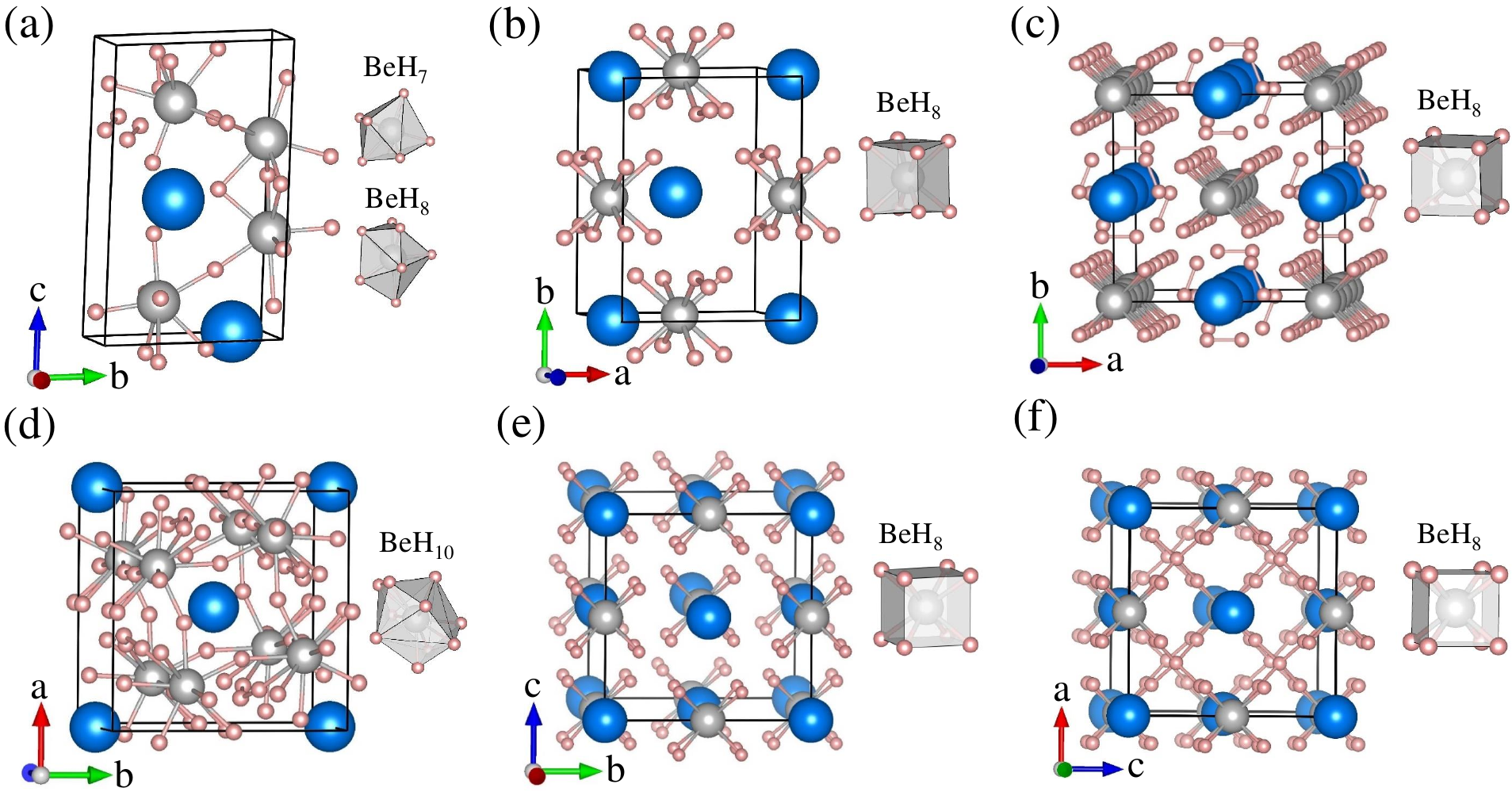}
  \caption{\label{fig:2} Structural configuration. The stable (a)--(c) and metastable structures (d)--(f) of Ac--Be--H compounds. (a) $P1$ AcBe$_2$H$_{10}$, (b) $Pmmn$ AcBeH$_8$, (c) $Cmcm$ AcBe$_2$H$_{14}$, (d) $P$4/$mbm$ AcBe$_2$H$_{16}$, (e) $Fm$$\bar{3}$$m$ AcBeH$_{8}$, and (f) $Fm$$\bar{3}$$m$ AcBeH$_{10}$ at 200~GPa. Ac atoms are depicted in blue, Be in gray, and H in pink.}
\end{figure*}

We discover three stable phases, $P1$ AcBe$_2$H$_{10}$, $Pmmn$ AcBeH$_8$, $Cmcm$ AcBe$_2$H$_{14}$ and four metastable phases, $Fm\bar{3}m$ AcBeH$_8$, $Fm\bar{3}m$ AcBeH$_{10}$, $C2/m$ AcBeH$_{12}$, and $P4/mbm$ AcBe$_2$H$_{16}$. These structures are all dynamically stable, metallic and superconducting. Our calculations predict that the $P$4/$mbm$ AcBe$_2$H$_{16}$ and $Fm$$\bar{3}$$m$ AcBeH$_{10}$ structures are superconductors with \Tc\ of 150~K at 200~GPa and 165~K at 300~GPa, respectively. Particularly interesting is a predicted metastable structure, \textit{fcc} AcBeH$_8$, that remains dynamically stable down to 10~GPa with a high \Tc\ of 181~K. Such a pressure is easy to achieve in an experiment, and is substantially lower than the required pressure to stabilize binary actinium hydrides.

\section{COMPUTATIONAL DETAILS}

The search for crystalline structures was performed using a
particle-swarm optimization algorithm, as implemented in the {\sc calypso}
code~\cite{wang2010crystal,wang2012calypso,gao2019interface,shao2022symmetry}. This
method has been extremely successful in predicting stable and
metastable superconducting hydrides~\cite{cui2019role}, some of which
have already been confirmed by experiments~\cite{flores2020perspective,zurek2019high,cui2020route,ma2021metal,shi2020formation}. The calculations to predict the lowest-enthalpy crystal structures of AcBe$_x$H$_y$ (with $x$ = 1-2, $y$ = 1-10, 12, 14, 16, 18, 20), considering up to four formula units, were done at 200~GPa. More than 2000 structures were sampled for each prediction run for every composition, and each generation of structures was evolved by selecting the 60\% lowest-enthalpy structures in the last step and randomly producing the remaining 40\%. The structure searches were considered converged when ~$\sim$ 1000 successive structures were generated without finding a new lowest-enthalpy structure. Structural relaxations and electronic band-structure calculations were performed using the projector augmented-wave (PAW) method as implemented in the Vienna $Ab$ $initio$ Simulation Package (VASP)~\cite{kresse1996efficient}. The exchange-correlation functional of density-functional theory was approximated by the generalized gradient approximation of Perdew, Burke, and Ernzerhof~\cite{perdew1996generalized}. The all-electron PAW method was adopted for Ac, Be, and H atoms with valence 6$s^2$6$p^6$6$d^1$7$s^2$, 1$s^2$2$s^2$, and 1$s^1$, respectively. The cutoff energy for the expansion of the wave function in the plane wave basis was set to 1000~eV. Monkhorst-Pack $k$-point meshes~\cite{tang2009grid} with a grid density of 0.20~\AA$^{-1}$ were chosen to ensure a total energy convergence better than 1~meV per atom. The phonon spectrum and electron-phonon coupling were calculated within linear-response theory with the {\sc quantum espresso} code~\cite{giannozzi2009quantum}. Ultrasoft pseudopotentials for Ac, Be and H were used  in EPC calculations~\cite{kresse1999ultrasoft}. The detailed encut, $k$-meshes and $q$-points for these seven compounds can be found in Table S1 of the SM~\cite{suppe} in the supplementary material. The superconducting critical temperatures are evaluated  based on the Allen-Dynes-modified McMillan equation~\cite{allen1975transition}. For value of the electron-phonon coupling constant $\lambda$ smaller than 1.5 we used the formula

\begin{equation}
  T_\text{c} = \frac{\omega_{\text{log}}}{1.2} \text{exp}\left[-\frac{1.04(1+\lambda)}{\lambda-\mu^*(1+0.62\lambda)}\right],
  \end{equation}
while in the very strong coupling regime ($\lambda>1.5$) we included the shape correction multipliers ($f_1$ and $f_2$) as follows:

\begin{equation}
 f_1 = [1 + (\frac{\lambda}{2.46(1+3.8\mu^*)})^{3/2}]^{1/3},
 \end{equation}

\begin{equation}
f_2 = 1 + \frac{(\frac{\bar{\omega}}{\omega_{\text{log}}}-1)\lambda^2}{\lambda^2+\left[1.82(1+6.3\mu^*)\frac{\bar{\omega}}{\omega_{\text{log}}}\right]^2}, \end{equation}

\begin{equation}
 T_\text{c}= f_1f_2\frac{\omega_{\text{log}}}{1.2}\text{exp}\left[\frac{-1.04(1+\lambda)}{\lambda-\mu^*(1+0.62\lambda)}\right], \end{equation}

where the logarithmic average frequency $\omega_{\text{log}}$ and mean square frequency $\bar{\omega}$ are defined as:
\begin{equation}
\omega_{\text{log}} =\text{exp}\left[\frac{\lambda}{2}\int\ln(\omega)\frac{\alpha^2F(\omega)}{\omega}d(\omega)\right],
  \end{equation}

  and
\begin{equation}
  \bar{\omega} = \sqrt{\frac{2}{\lambda}\int\alpha^2F(\omega)\omega d(\omega)},
  \end{equation}
and where $\alpha^2F(\omega)$ is the Eliashberg spectral function.

\section{RESULTS AND DISCUSSION}

We calculated the phase diagram of AcBe$_x$H$_y$ at 200~GPa. The results are shown in Fig.~\ref{fig:1} and Figs.~S1(a) and (b). Three ternary phases with stoichiometries AcBe$_2$H$_{10}$, AcBeH$_8$, and AcBe$_2$H$_{14}$ become stable against decomposition into elemental or binary solids. Additionally, we uncovered several metastable structures with distances to the convex hull smaller than around 10~meV/atom, namely, AcBeH$_6$ (9.3~meV/atom), AcBe$_2$H$_8$ (2.8~meV/atom), AcBe$_2$H$_{12}$ (2.4~meV/atom), AcBe$_2$H$_{16}$ (5.0~meV/atom), and AcBeH$_{12}$ (10.9~meV/atom). In view of the fact that hydrogen-rich phases with high symmetry have an increased potential to exhibit high-\Tc\ superconductivity~\cite{xie2020hydrogen, peng2017hydrogen,du2022room,sun2022prediction,hou2022superconductivity,jiang2022ternary}, we also consider two metastable phases with \textit{fcc} symmetry AcBeH$_8$ (4~meV/atom) and AcBeH$_{10}$ (35~meV/atom). Focusing for the AcBeH$_8$ phase at a distance of only 4~meV/atom to the convex hull, we compared relative enthalpies with respect to competing structures and possible decomposition products from 0 to 500~GPa, as shown in Figs.~S1(c) and (d). In comparison to the work of Wan et al.~\cite{wan2022superconductivity}, we performed prediction runs for several compositions of Ac--Be--H phase diagram and not only for XBeH$_8$. Fig.~S1(a) indicates that the formation enthalpy of \textit{fcc} phase of AcBeH$_8$ is higher than the one of the thermodynamically stable $Pmmn$ phase and of other metastable structures, as shown in Fig.~S1(d), but all the decomposition products exhibit positive formation energies relative to \textit{fcc} AcBeH$_8$ above 30~GPa. The structure of AcBeH$_8$ with symmetry $Fm\bar{3}m$ is not thermodynamically stable in the range of pressures 0--30~GPa but there is a great possibility to synthesize this structures experimentally at higher pressures. Because the pressure is extremely low, it is worthy of experimental exploration. Other low-enthalpy structures have a reasonable possibility to be synthesized experimentally thanks to their small enthalpy distance to the convex hull at 200~GPa~\cite{kostrzewa2020lah,peng2017hydrogen,drozdov2019superconductivity,somayazulu2019evidence,semenok2021superconductivity,chen2021high,salke2019synthesis}.

\begin{figure*}[htp]
\centering
  \includegraphics[width=1\linewidth,angle=0]{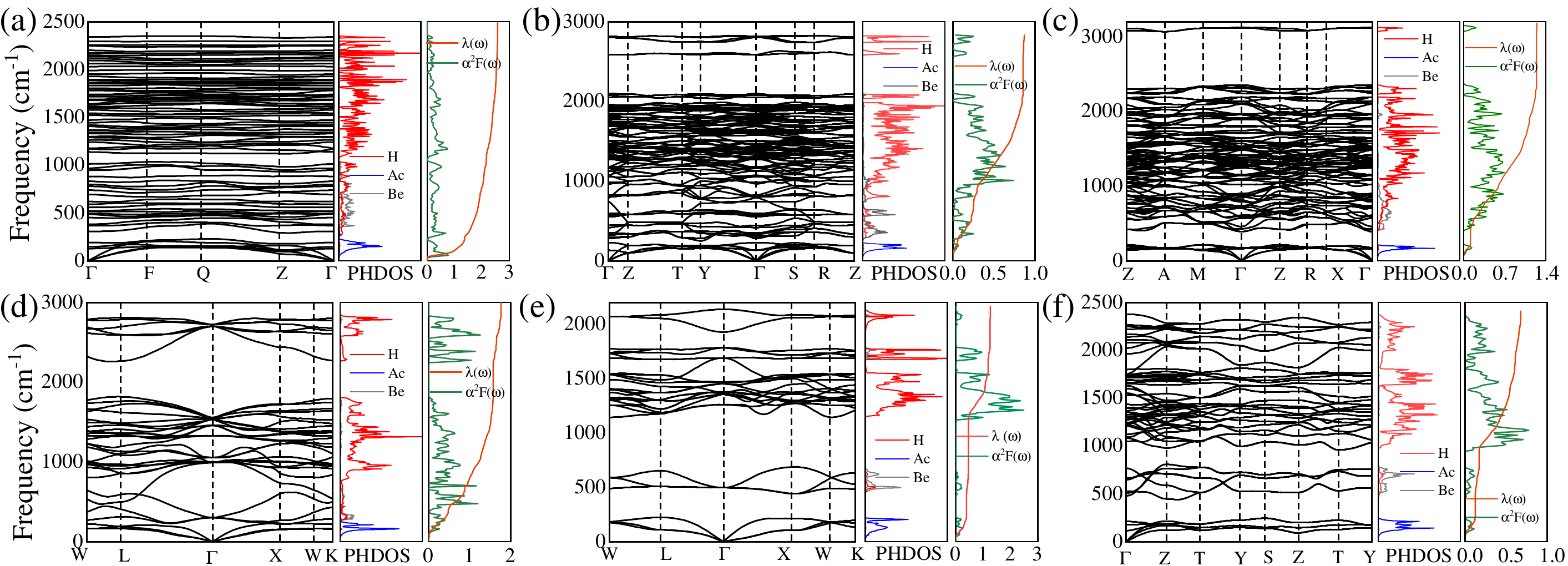}
  \caption{\label{fig:3} Phonon dispersion and electron-phonon coupling (EPC) coefficient of $\lambda$($\omega$). Phonon dispersion, the electron-phonon coefficient $\lambda$($\omega$), and the Eliashberg spectral function $\alpha^2F(\omega)$ of (a) $P1$ AcBe$_2$H$_{10}$, (b) $Pmmn$ AcBeH$_8$, (c) $Cmcm$ AcBe$_2$H$_{14}$, (d) $P$4/$mbm$ AcBe$_2$H$_{16}$, (e) $Fm\bar{3}m$ AcBeH$_{8}$ at 200~GPa, and (f) $Fm\bar{3}m$ AcBeH$_{10}$ at 300~GPa.}
\end{figure*}

The crystal structures of thermodynamically stable and metastable AcBe$_x$H$_y$ compounds are shown in Figs.~\ref{fig:2} and~S2. AcBe$_2$H$_{10}$ adopts a low-symmetry $P1$ structure [Fig.~\ref{fig:2}(a)] that contains BeH$_8$, BeH$_7$ units, and H$_2$ molecules, characterized by Be--H and H-H distances of 1.29--1.59~\AA~and 0.87~\AA, respectively. The orthorhombic phase of AcBeH$_8$ [Fig.~\ref{fig:2}(b)] with space group $Pmmn$ is composed of BeH$_8$ octahedra with a Be--H distance of 1.30--1.43~\AA~, as well as Ac atoms occupying the $2a$ Wyckoff positions. AcBe$_2$H$_{14}$ has is also orthorhombic with $Cmcm$ symmetry [Fig.~\ref{fig:2}(c)]. In this structure, each Ac occupies the 4$b$ Wyckoff positions and three pairs of H$_2$ molecules, with bond lengths of around 0.85~\AA, are positioned between pairs of Ac atoms. Additionally, two inequivalent Be atoms occupy the 4$a$ and 4$c$ Wyckoff positions, respectively. These Be atoms are surrounded by eight H atoms forming BeH$_8$ octahedra with Be--H distances of 1.36--1.56~\AA~at 200~GPa. In the metastable structure of AcBe$_2$H$_{16}$ with $P4/mbm$ symmetry [Fig.~\ref{fig:2}(d)], the Ac atoms occupy the 4$g$ position: each Be is bonded with ten H atoms to form a hexadecahedal BeH$_{10}$ unit. Adjacent BeH$_{10}$ units along the $y$ axis are connected with each other by sharing one hydrogen atom, and neighbouring BeH$_{10}$ units along the $c$ axis are connected by H$_2$ and H$_3$ units with H-H distances of 1.08 and 0.92~\AA~at 200~GPa, respectively. The covalent nature of these latter bonds can be confirmed by analysis of the electron localization function (ELF), as shown in Fig. S3(a). The \textit{fcc} phase of AcBeH$_8$ [Fig.~\ref{fig:2}(e)], isostructural to $Fm$$\bar{3}$$m$ LaB(Be)H$_8$~\cite{liang2021prediction,di2021bh,zhang2022design}, consists of BeH$_8$ hexadra that occupy the octahedral sites of the \textit{fcc} lattice formed by Ac atoms. The AcBeH$_{10}$ phase with $Fm\bar3m$ symmetry  [Fig.~\ref{fig:2}(f)] is 15~meV/atom energetically higher than the lowest enthalpy phase $P2_1$. Compared with \textit{fcc} AcBeH$_8$, the four additional H atoms of AcBeH$_{10}$ occupy tetrahedral sites that are connected by four BeH$_8$ octahedra to form H$_5$ regular tetrahedra with H-H distances of 1.00~\AA. Other metastable structures (e.g., $P$1 AcBeH$_6$, $P$-1 AcBe$_2$H$_8$, $Cmmm$ AcBe$_2$H$_{12}$ and $C2/m$ AcBeH$_{12}$) are shown in Fig. S2. In these structures, the Be and Ac atoms donate electrons to hydrogen, forming typical ionic hydrides (see Table~SII). Apart from the Be--H bonds, these metastable structures present also H-kagome lattices, together with H$_2$ and H$_3$ units (see Fig.~S2).

\begin{figure*}[htp]
\centering
  \includegraphics[width=0.9\linewidth,angle=0]{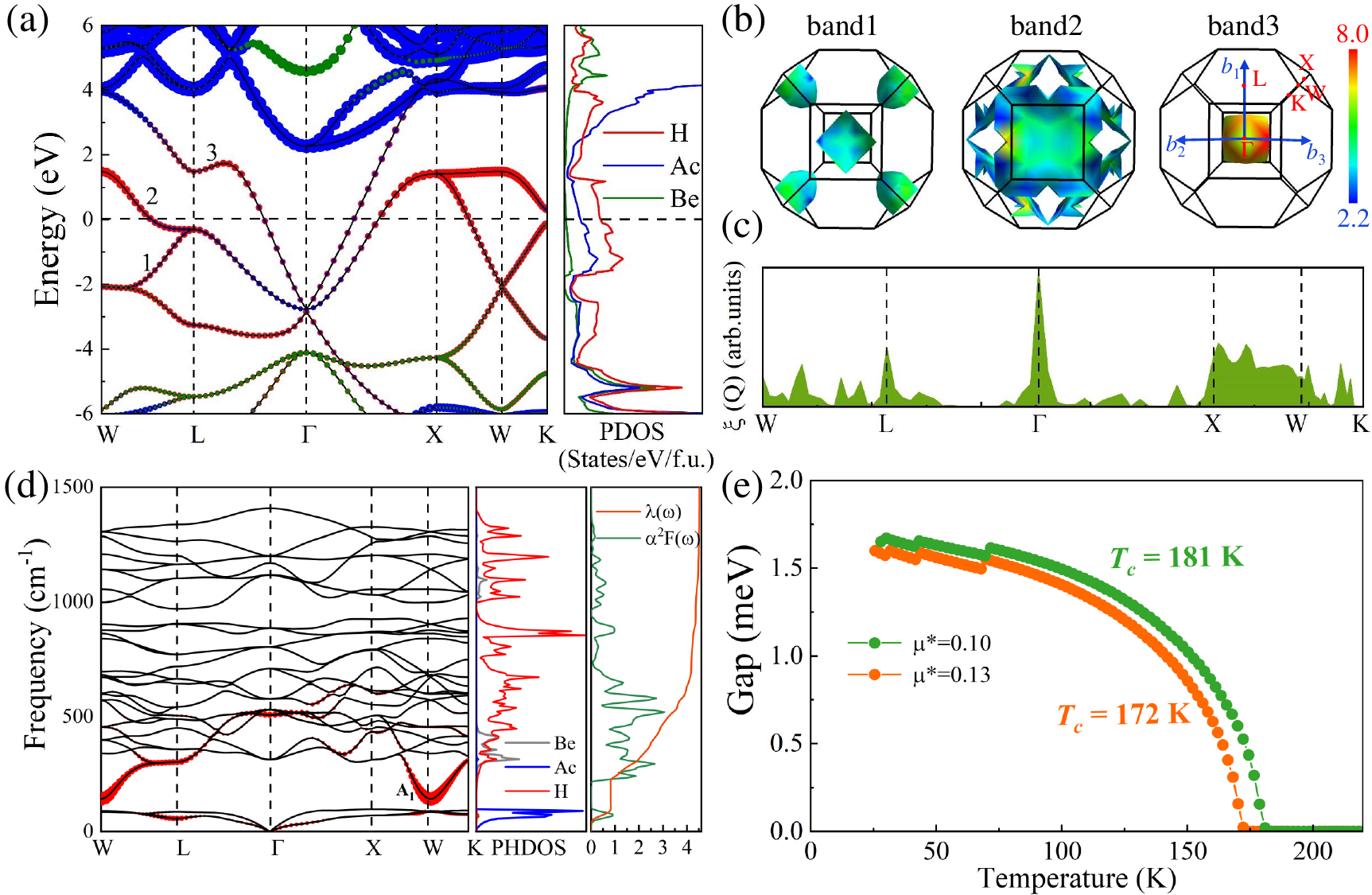}
  \caption{\label{fig:4}Electronic and superconductivity properties of AcBeH$_8$ at 10~GPa. Calculated (a) band structures and projected density of states (PDOS), (b) Fermi surfaces corresponding to the three bands crossing the Fermi energy level, colored with respect to the Fermi velocity $\langle$v$\rangle$ (10$^5$~m$/$s), (c) nesting function $\xi$($Q$) along some special $Q$ trajectories, (d) phonon dispersions, the EPC coefficient of $\lambda$($\omega$) and Eliashberg spectral function $\alpha^2F(\omega)$, and (e) anisotropic superconducting gap of $Fm\bar{3}m$ AcBeH$_{8}$ at 10~GPa. Red solid circles represent the phonon linewidth with the radius proportional to the respective coupling strength.}
\end{figure*}

In order to further explore the fascinating properties of Ac--Be--H compounds, we calculated their electronic structure and phonon properties. Almost all predicted structures are also dynamically stable, as confirmed by inspection of the phonon bands, with the exception of AcBe$_2$H$_{12}$ and AcBeH$_{10}$ that display imaginary phonons (see Figs.~\ref{fig:3} and S4). Also AcBe$_2$H$_{12}$ and AcBeH$_{10}$ [Figs. S4(a) and (d)] are not dynamically stable at 200~GPa, however, AcBeH$_{10}$ becomes stable at higher pressure [Fig.~\ref{fig:3}(f)]. The case of  $Fm$$\bar{3}$$m$ AcBeH$_8$ is particularly interesting as it remains dynamically stable down to pressures of 10~GPa, as shown in Figs.~\ref{fig:4}(d) and S5. All structures are metallic with several bands crossing the Fermi energy $E_F$, and have therefore promising electronic bands for high-temperature superconductivity [see Figs.~S6 and S7]. The significant overlap of the partial electronic density of states (DOS) of the different atoms indicates a strong hybridization of Ac--H and Be--H under pressure. The results clearly indicate that hydrogen atoms make a substantial contribution to the total DOS near $E_F$, e.g., for $Cmcm$ AcBe$_2$H$_{14}$ (49\%), $P$4/$mbm$ AcBe$_2$H$_{16}$ (52\%), $C$2/$m$ AcBeH$_{12}$ (49\%), and $Fm$$\bar{3}$$m$ AcBeH$_{8}$ (44\%) at 200~GPa (Table~SIII).

\begin{figure}[htp]
\centering
  \includegraphics[width=0.95\linewidth,angle=0]{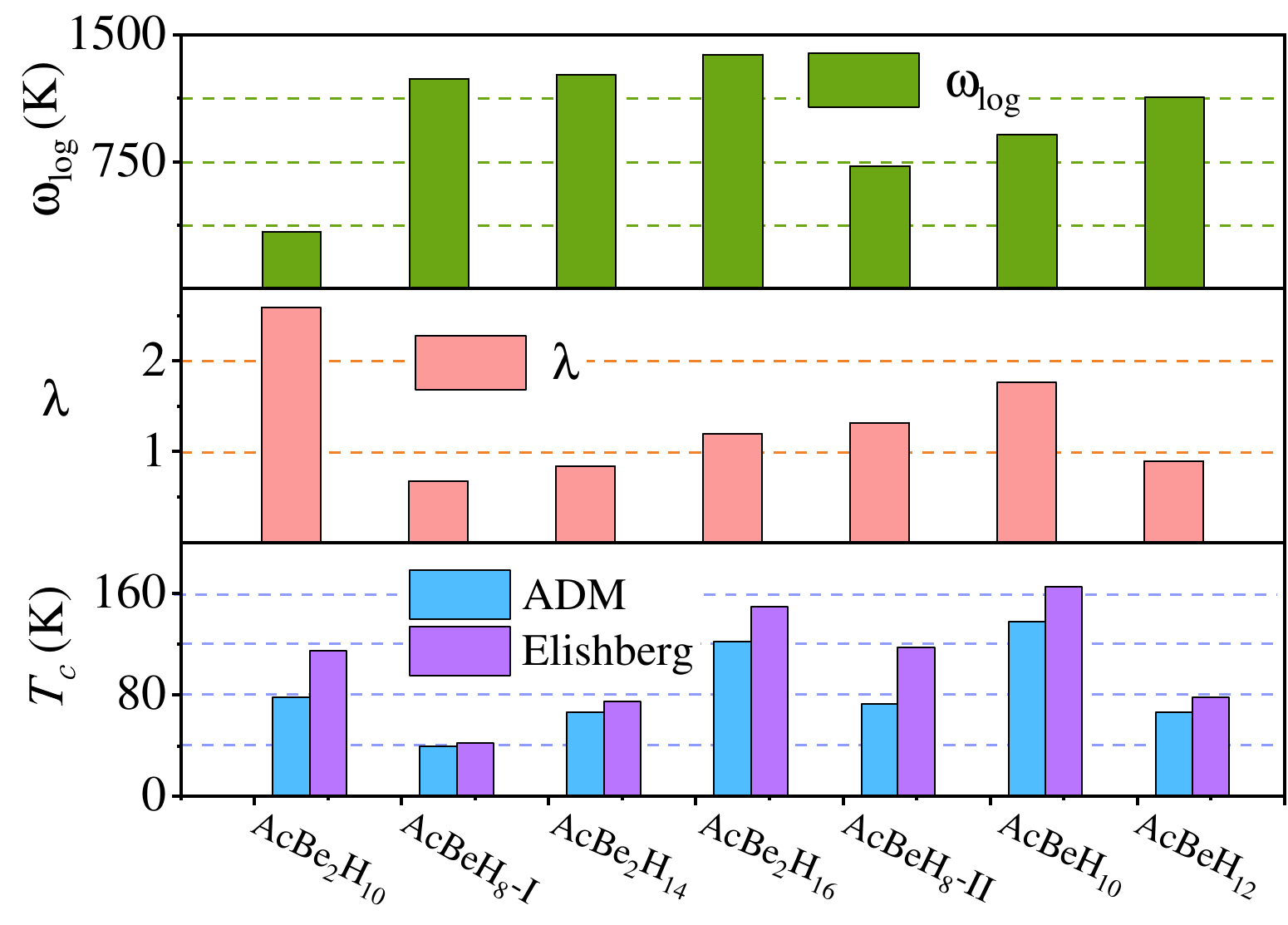}
  \caption{\label{fig:5} Superconducting-transition temperature. Logarithmic average phonon frequency ($\omega$$_{\text{log}}$), EPC parameter $\lambda$, and \Tc\ ($\mu^*$ = 0.10)~for $P1$ AcBe$_2$H$_{10}$, $Pmmn$ AcBeH$_8$ (I), $Cmcm$ AcBe$_2$H$_{14}$, $P$4/$mbm$ AcBe$_2$H$_{16}$, $Fm\bar{3}m$ AcBeH$_{8}$ (II), $C$2/$m$ AcBe$_2$H$_{12}$ at 200~GPa, and $Fm\bar{3}m$ AcBeH$_{10}$ at 300~GPa.}
\end{figure}

The contribution of hydrogen to the DOS at $E_F$ is large for the three stable structures $P1$ AcBe$_2$H$_{10}$, $Pmmn$ AcBeH$_8$, and $Cmcm$ AcBe$_2$H$_{14}$, with values 0.147, 0.141 and 0.130~eV$^{-1}$ at 200~GPa, respectively [Figs. S6(a)-(c)]. For the highly symmetric metastable structures, this contribution is even larger, especially in the case of $Fm$$\bar{3}$$m$ AcBeH$_8$, where we can observe an enhancement from 44\% (200~GPa) to 65\% (10~GPa). The corresponding contributions to the DOS are 0.215 and 0.261~eV$^{-1}$, respectively. This facilitates the formation of more Cooper pairs at $E_F$, thereby increasing the superconducting temperature \Tc\ [Figs.~\ref{fig:4}(a) and S6(e)]. There are three bands of~\textit{fcc} AcBeH$_8$ at 10~GPa that cross $E$$_F$, yielding the Fermi surface (FS) shown in Figs.~\ref{fig:4}(a) and (b). Bands 1 and 2 are degenerate along the high-symmetry line from $L$ to $X$ in the first Brillouin zone and their FS exhibits pockets at $X$ and complex semi-closed shapes with a hexagonal opening, respectively. In addition to this, band 3 forms a closed electron-like surface at the $\Gamma$ point, which leads to larger DOS at $E_F$ and to a further enhancement of the EPC~\cite{simon1997superconductivity}. Despite the higher hydrogen content, the contribution of the H atoms to the DOS of $Fm$$\bar{3}$$m$ AcBeH$_{10}$ at  $E_F$ is less than that of BeAcH$_8$ [only 0.140 eV$^{-1}$ at 300~GPa, see Fig. S6(f)], due to the extra hydrogen molecules located at deep energy levels. For other metastable structures with less symmetry (e.g., $P$1 AcBeH$_6$, $P$-1 AcBe$_2$H$_8$, and $C2/m$ AcBeH$_{12}$), the partial DOS of H at $E$$_F$ amounts to less than $\sim$1~eV$^{-1}$, leading to reduced superconducting transition temperatures.

To investigate superconductivity and its mechanism, we calculated phonon dispersion curves, partial phonon density of states (PHDOS), the Eliashberg spectral function $\alpha^2F(\omega)$ and the EPC integrated $\lambda$($\omega$) of Ac--Be--H hydrides and we plotted these quantities  in Figs.~\ref{fig:3} and S5. The PHDOS of all structures has common features: the heaviest Ac atoms contribute to the lowest-frequency vibrational modes, the middle-frequency modes are mainly dominated by Be and H vibrations, and the high-frequency modes are exclusively due to H atoms.

\begin{table}
\centering
\caption{Superconducting paraments of AcBeH$_8$. The calculated EPC parameter ($\lambda$), logarithmic average phonon frequency ($\omega$$_{\text{log}}$), and the estimated \Tc\ for selected structures using the  Allen-Dynes modified McMillan (ADM) equation~\cite{allen1975transition}, and numerically solving the Eliashberg equations~\cite{eliashberg1960interactions} with $\mu^* = 0.10$ for $Fm\bar{3}m$ AcBeH$_{8}$ at high pressures.}
\label{Cpa}
\renewcommand{\arraystretch}{2}

\begin{tabular}{ccccc}
\hline
\hline
                                &             &                 & \multicolumn{2}{c}{\textbf{\Tc\ (K)}} \\
 \cline{4-5}

 ~\textbf{Pressure (GPa)}      &~\textbf{$\lambda$}      &~\textbf{$\omega_{\text{log}}$ (K)}     &~\textbf{ADM}  &~\textbf{Eliashberg} \\
\hline
            200                 &1.32            &724                &73     &118  \\
            150                 &1.33            &876                &89     &132  \\
            100                 &1.47            &846                &115    &141  \\
            70                  &1.63            &835                &117    &155  \\
            30                  &2.26            &706                &123    &170   \\
            10                  &4.50            &433                &129    &181    \\
\hline
\hline
\end{tabular}
\end{table}

The superconducting properties of the Ac--Be--H system can be estimated using the McMillan formula (as modified by Allen-Dynes~\cite{allen1975transition}) or by solving numerically the Eliashberg equations~\cite{eliashberg1960interactions} We used the typical value of the Coulomb pseudopotential $\mu^*$= 0.10. A summary of the results can be found in Fig.~\ref{fig:5}.
Among the thermodynamically stable compounds, AcBe$_2$H$_{10}$ possesses the strongest EPC parameter $\lambda = 2.59$, mainly owing to the soft modes along the $Z$-$\Gamma$ high-symmetry line [Fig.~\ref{fig:3}(a)], yielding a \Tc\ of 115~K at 200~GPa. This is the highest \Tc\ among the thermodynamically stable compounds [see Fig. S8(a)]. The low frequency region (0--245~cm$^{-1}$), dominated by the Ac atoms, contributes a significant amount ($\sim$ 53\% ) to the total $\lambda$, but also the range of intermediate frequencies (245--1063~cm$^{-1}$, related to the Be--H bonds and the high-frequency region (1081--2358~cm$^{-1}$, due to H atoms only) contribute 30\% and 17\%, respectively, to $\lambda$, as shown in Table SIV.
The two other stable compounds $Pmmn$ AcBeH$_8$ and $Cmcm$ AcBe$_2$H$_{14}$ have values of $\lambda$ of 0.67 and 0.84, respectively at 200~GPa, and in both cases H atoms  give the largest contribution to $\lambda$ with 74\% and 54\%, respectively [Figs.~\ref{fig:3}(b)-(c) and Table SIV]. The metastable structures AcBe$_2$H$_{16}$ and AcBe$_2$H$_{12}$ have values of $\lambda$ of 1.19 and 0.9 at 200 GPa, yielding a \Tc\ of 150~K and 78~K, respectively (Figs.~\ref{fig:5} and S9).

In what concerns \textit{fcc} AcBeH$_{10}$, we remark that it is not stable at 200~GPa [Fig. S4(a)] is stabilized at the higher pressure 300~GPa [Fig.~\ref{fig:3}(f)]. EPC calculations show that $Fm\bar{3}m$ AcBeH$_{10}$ phases are promising conventional superconductors with $\lambda=1.76$ and $\omega_{\text{log}}=909$~K at 300~GPa. This leads to an estimated \Tc\ of 165~K [Figs.~\ref{fig:5} and S8(e)]. The phonon modes at frequencies between 427 and 2858~cm$^{-1}$, corresponding to the hydrogen atoms, contribute a large amount (72\%) to the total $\lambda$.

The other \textit{fcc} structure, AcBeH$_8$, is dynamically stable from 200 to 10~GPa, as shown by the phonon dispersion depicted in Figs.~\ref{fig:3}(e), S5 and ~\ref{fig:4}(d). The calculated $\lambda$ and phonon frequency logarithmic average $\omega_{\text{log}}$ are 1.32 and 724~K at 200~GPa, respectively, leading to a \Tc\ of 118~K for $\mu^*= 0.10$. Note that the vibrations related to H atoms at high frequencies give a significant contribution to the total $\lambda$ (64--78\%) for pressures in the range of 30--200~GPa, and the contribution of Be and H at intermediate frequencies only accounts for less than 6\%. As the pressure decreases, the critical temperature of AcBeH$_8$ is gradually enhanced (see Tables I and SIV). Strikingly, $\lambda$ increases to 4.5 with a \Tc\ of 181~K at 10~GPa [see Figs.~\ref{fig:4}(d)-(e)]. We cannot compare directly with the \Tc\ reported in Ref.~\cite{wan2022superconductivity}, 284~K at 150~GPa, as we used the McMillan formula (as modified by Allen-Dynes~\cite{allen1975transition}) and solved numerically the Eliashberg equations~\cite{eliashberg1960interactions} instead of EPW to calculate the superconducting transition temperature. High temperature superconductivity is prone to appear when $\lambda$ is large, as it happens, e.g., for CSH$_7$ ($\lambda$ = 3.06, \Tc\ = 170~K@150~GPa)~\cite{cui2020route}, LaH$_{10}$($\lambda$ = 3.4, \Tc\ = 274~K@210~GPa)~\cite{liu2017potential}, YH$_9$ ($\lambda$ = 4.42, \Tc\ = 276~K@150~GPa)~\cite{peng2017hydrogen}, ScLuH$_{12}$ ($\lambda$ = 4.43, \Tc\ = 266~K@100~GPa), Y$_3$YbH$_{24}$ ($\lambda$ = 4.78, \Tc\ = 222~K@100~GPa), YLu$_3$H$_(24)$ ($\lambda$ = 4.78, \Tc\ = 288~K@110~GPa), and CaLu$_2$H$_{18}$ ($\lambda$ = 4.13, \Tc\ = 299~K@140~GPa)~\cite{du2022room}. When the pressure decreases progressively to 10~GPa, the interaction between Be and H becomes stronger, and the contribution to $\lambda$ due to soft phonon-modes (110--540~cm$^{-1}$) increases to 60\%, especially thanks to contributions from the $W$ point with symmetry $C_{2v}$, originating from $A_1$ modes that are mainly related to vibrations of hydrogen atoms [Figs.~\ref{fig:4}(d) and S10]. There also appears a significant increase in the nesting function at 10~GPa along $W$-$L$ and $X$-$W$-$K$, which is perfectly consistent with the phonon softening at $W$, as shown in the nesting function $\xi$($Q$) [Figs. \ref{fig:4}(c) and  S11]. Our calculations indicate that under a pressure larger than 10~GPa, the main contribution to superconductivity comes from the H phonon modes, but down to very low pressure, the interaction between Be and H plays a key role in enhancing superconductivity while ensuring structural stability.

\section{CONCLUSIONS}

In summary, we have investigated the crystal structures and superconductivity of ternary Ac--Be--H systems at 200~GPa combining crystal structure prediction and first-principle calculations. We uncover  three  thermodynamically stable compounds with stoichiometries
AcBe$_2$H$_{10}$, AcBeH$_8$, AcBe$_2$H$_{14}$, as well as four metastable superconducting compounds AcBe$_2$H$_{16}$, \textit{fcc} AcBeH$_{10}$, \textit{fcc} AcBeH$_8$ and AcBe$_2$H$_{12}$. All of these structures exhibit metallic nature. Electron-phonon coupling calculation shows that AcBe$_2$H$_{16}$ and \textit{fcc} AcBeH$_{10}$ are good phonon-mediated superconductors, with \Tc\ of 150~K at 200~GPa and 165~K at 300~GPa, respectively. More interestingly,~\textit{fcc}-AcBeH$_8$ remains dynamically stable down to 10~GPa where it exhibits a \Tc\ of 181~K.
The soft phonon vibration modes originating from Be--H interactions in BeH$_8$ units contribute to the enhancement of superconductivity with decreasing pressure. We expect that our results will stimulate more research on ternary superconducting hydrides with high critical temperature and stable at extremely low pressure.

\section{ACKNOWLEDGMENTS}

The authors acknowledge funding from the NSFC under grants No. 12074154, No. 11804128, No. 12174160, and No. 11804129, and No. 11722433. W.C. and M.A.L.M. acknowledge the funding from the Sino-German Mobility Programme under No. M-0362. Y.L. acknowledges the funding from the Six Talent Peaks Project and 333 High-level Talents Project of Jiangsu Province. A.P.D. is grateful for financial support from the National Science Centre (Poland) through Project No. 2022/47/B/ST3/00622.
K.G. acknowledges financial support from the China Scholarship Council. All the calculations were performed at the High Performance Computing Center of the School of Physics and Electronic Engineering of Jiangsu Normal University.

\bibliography{AcBeH}
\end{document}